\documentclass{LT23auth}
\usepackage{graphicx}
\usepackage[]{amsmath}
\usepackage[]{latexsym}

\begin{document}

\begin{frontmatter}

\title{
Dirac Monopole and Spin Hall Conductance for Anisotropic
Superconductivities 
}

\author[address1]{Yasuhiro Hatsugai \thanksref{thank1}},
\author[address2]{Shinsei Ryu }

\address[address1]{
Dept. of Applied Physics Univ. of Tokyo 7-3-1 Hongo,
Bunkyo-ku, Tokyo, JAPAN and\\
PRESTO, JST, Saitama 332-0012, JAPAN
}

\address[address2]{Dept. of Applied Physics Univ. of Tokyo, 7-3-1 Hongo,
Bunkyo-ku, Tokyo, JAPAN
}

\thanks[thank1]{ Corresponding author. E-mail:hatsugai@pothos.t.u-tokyo.ac.jp\hfil}

\begin{abstract}
Concept of the topological order is useful to characterize 
anisotropic superconductivities.
The spin Hall conductance
 distinguishes superconductivities with the same symmetry.
The Chern number for the spin Hall conductance is given by the 
covering degree of a closed surface around the Dirac monopole.
 It gives a clear illustration for the on-critical condition for the
$d_{x^2-y^2}$ supreconductivitiy with non zero chemical potential.
Non trivial topological orders on the  triangular lattice are also
presented and demonstrated by the topological objects.
\end{abstract}

%
%
\end{frontmatter}


\newcommand{\seceq}{  (\thesection - \theequation)}
\newcommand{\mb}[1]{\mbox{\boldmath $#1$}}
\newcommand{\aeq}[0]{&=&}
\newcommand{\es}[0]{\[}
\newcommand{\ee}[0]{\]}
\newcommand{\eas}[0]{\begin{eqnarray*}}
\newcommand{\eae}[0]{\end{eqnarray*}}
\newcommand{\les}[0]{\begin{equation}}
\newcommand{\lee}[0]{\end{equation}}
\newcommand{\leas}[0]{\begin{eqnarray}}
\newcommand{\leae}[0]{\end{eqnarray}}
\newcommand{\mchs}[2]
{
\left\{
\begin{array}{l}
#1    \\
#2
\end{array}
\right.
}
\newcommand{\mchss}[4]
{
\left\{
\begin{array}{cc}
#1 & #2   \\
#3 & #4
\end{array}
\right.
}
\newcommand{\mat}[4]
{
\left(
\begin{array}{cc}
#1 & #2 \\
#3 & #4 
\end{array}
\right)
}
\newcommand{\mvec}[2]
{
\left(
\begin{array}{c}
#1  \\
#2  
\end{array}
\right)
}
\newcommand{\mvecthree}[3]
{
\left(
\begin{array}{c}
#1  \\
#2  \\
#3  
\end{array}
\right)
}
\newcommand{\mvecfour}[4]
{
\left(
\begin{array}{c}
#1  \\
#2  \\
#3  \\
#4  
\end{array}
\right)
}
\newcommand{\mvecfive}[5]
{
\left(
\begin{array}{c}
#1  \\
#2  \\
#3  \\
#4  \\
#5  
\end{array}
\right)
}
\newcommand{\mvecsix}[6]
{
\left(
\begin{array}{c}
#1  \\
#2  \\
#3  \\
#4  \\
#5  \\
#6  
\end{array}
\right)
}
\newcommand{\minv}[1]{#1^{-1}}
\newcommand{\minfty}[0]{{-\infty}}
\newcommand{\cosec}[0]{{\rm \ cosec}}
\newcommand{\im}[0]{{\rm Im\ }}

\newcommand{\mvs}[0]{\vskip 0.5cm}


Topological quantum phase transition is conceptually 
interesting since it is a zero temperature phase transition
without symmetry breaking\cite{wen_topological}.
 It is associated with a change of topological characterization 
for the quantum mechanical ground state. 
One of the widely established examples is a quantum Hall plateau
transition between different Hall states.
Symmetries are the same 
but only the Hall conductances are different among them.
The Hall conductance has an intrinsic 
topological character and the
plateau transition is specified by the change of such  topological  objects.
Typical realizations of them are  the Chern numbers, 
vortices and  edges states\cite{tknn,yh}.
Also special form of a selection rule is expected  and 
the possible stability of the phase 
which puts special importance on the topological transitions 
discriminates from other quantum phase transitions.

Recently, there have been trials to extend the concept
of this topological phase transition to unconventional singlet 
superconductivity.\cite{smf,ym-yh}
The superconducting system is  mapped  into the  Hall
system and analogy between them is used to characterize the
superconducting ground state.
Here the ``spin'' Hall conductance plays a main role
which is given by the Chern numbers. 
Recently proposed superconductivity with time-reversal symmetry breaking 
is a non trivial example with the topological order.
\cite{volovik1,laughlin-dd,smf,ym-yh}.
 
The topological phase transition in 
superconductivity raises important theoretical questions 
how the topological character restricts transition types.
 Starting from the  generalized lattice Bogoliuvov-de Gennes (B-dG)
hamiltonian for superconductivity, we discuss the
topological quantum phase transition and demonstrate the topological
character\cite{yh-sr}.
 In this paper how the gap opening and the topological objects are 
related is focused. 
Extension to the triangular lattice
which is the simplest system with frustration is also discussed.

We start from the following lattice 
B-dG hamiltonian for superconducting quasi-particles
$
H
= 
 \sum_{ij} (
t_{ij}
 c_{i\sigma}^\dagger c_{j\sigma}
 + 
\Delta_{ij} 
c_{i\uparrow}^\dagger c_{j\downarrow}^\dagger
+ 
\Delta_{ij} ^*
c_{j\downarrow} c_{i\uparrow})
-\mu \sum_{i\sigma} c_{i\sigma}^\dagger c_{i\sigma} $. 
We assume the system is translational invariant as
 $t_{ij}=t_{i-j}$, $\Delta_{ij}=\Delta_{i-j}$
and  further require $t_{ij}=t_{ji}$ ( $\epsilon(\mb{k})=\epsilon(-\mb{k})$).
Then $
H=   \sum_k 
\mb  {c}^\dagger (\mb{k} ) \mb{h} (\mb{k} )\mb  {c} $
with
$
\mb{h} (\mb{k} ) =  
 \mb{R} (\mb{k})  \cdot \mb{\sigma}  
$ 
where 
$\mb{\sigma}=(\sigma_x,\sigma_y,\sigma_z)$  are  the Pauli
matrices 
and 
$ \mb{R} = \mb{R}(\mb{k} ) =
 (R_x,R_y,R_z)=
(
{\rm Re\, }\Delta(\mb{k}),
-{\rm Im\, }\Delta(\mb{k}),
\epsilon(\mb{k}) )
$, 
where 
$
\mb  {c}^\dagger (\mb{k} )
=
(c^\dagger _\uparrow(\mb{k}),c_\downarrow(\mb{k}) )
$,
$\epsilon( \mb{k}) 
= 
\sum_j  e^{-i \mb{k} \cdot \mb{r}_j} t_j-\mu $ and
$\Delta( \mb{k}) 
= 
\sum_j  e^{-i \mb{k} \cdot \mb{r}_j} \Delta_j
$.
The ``spin'' Hall conductance of the 
superconducting state on a lattice is given by
 the generalized Thouless-Kohmoto-Nightingale-den Nijs (TKNN) formula 
as
$
\sigma = -\frac {e^2}{h} C
$ 
$=   \frac {1}{2\pi i} \int_{T^2} d \mb{S}_k\,\cdot
{\rm rot } _k \,   \mb{A} _k 
$
$
= \int _{T^2}
  dk_x \wedge dk_y 
(
\langle \partial_x  \mb{k}  |\partial_y  \mb{k}  \rangle  
-
\langle \partial_y \mb{k}  |\partial_x   \mb{k} \rangle  
)
$
where $
\mb{A}_k=  \langle \mb{k}  | \mb{\nabla }_k
 \mb{k} \rangle 
$,
 $ \mb{h}(\mb{k} ) |  \mb{k} \rangle  =  -E(\mb{k} )|\mb{k} \rangle $,
$E(\mb{k} )=\sqrt{\epsilon(\mb{k} )^2+| \Delta (\mb{k} )|^2  }$
\cite{tknn,yh,ym-yh}.
This expression for the Chern number is rewritten
in $\mb{R}$ space as
$
C = 
\frac {1}{2\pi i} \int_{R(T^2)} d \mb{S}_R\,\cdot
{\rm rot } _R \,  \mb{A} _R
$
where
$\mb{A}_R = 
   \langle \mb{R} | \mb{\nabla }_R
\mb{R}\rangle $. \cite{yh-sr}
In a particular gauge,
$
| \mb{R}\rangle  = 
\, ^t\!(
{-\sin \frac { \theta }{2} }
{e^{i\phi}\cos \frac { \theta }{2} })
$
and 
$\mb{A}_R
=
i \frac {\sin\theta }{2R(1-\cos\theta )} \hat e_\phi
$
where $(R,\theta,\phi)$ is a polar coordinate of $\mb{R} $.
The integral region $R(T^2)$ is 
a closed surface in three dimensional parameter space
mapped from the Brillouin zone $T^2$.
The corresponding vector potential defines a magnetic monopole
at the origin as
$
{\rm div}\,  {\rm rot}\,  \mb{A} _R =  - {2\pi i} \, \delta_R(\mb{R} )
$ \cite{berry}.
Therefore, by the Gauss' theorem, 
we have another expression for the Chern number as
\begin{eqnarray*} 
C &=&  - \int_{R(T^2 )} dV  \delta_R(\mb{R} )= -N(R(T^2),O)= -N_{covering}
\end{eqnarray*} 
where $N_{covering}=N(R(T^2),O) $ is a degree of covering by the closed
surface
$R(T^2)$ around the origin $O$ \cite{yh-sr}.

We use the above topological expression to
discuss the gap closing ( on-critical ) condition for the order
parameters.
We assume a form of the order parameter used in ref\cite{yh-sr}.
It tells 
$\epsilon(\mb{k})=-2 t (\cos k_x+\cos k_y)-\mu$,
 $
\Delta (\mb{k} ) = \Delta_0+
2 \Delta_{x^2-y^2} ( \cos k_x-\cos k_y)
+
2i \Delta_{xy} (\cos (k_x+k_y)-\cos(k_x- k_y))
$.
In the singlet case, 
the monopole, is doubly covered since 
$ \mb{R} (\mb{k} )= \mb{R} (-\mb{k} )$ 
which implies the selection rule $\Delta C = \pm 2$\cite{ym-yh}.
One of the  interesting cases is given by the $\Delta_{xy}=0$. 
In this situation, the energy gap collapses. 
Then due to the gapless Dirac dispersion,
one can not determine the Chern number 
without ambiguity ( when $|\mu|$ is small). 
This situation is clear by the present demonstration since
the surface $R(T^2)$ is collapsed into a diamond shaped two dimensional
region 
( See Fig.1). 
When the Dirac monopole $O$ is on this region, 
the Chern number is ill-defined since the dispersion
is gapless. This condition is clearly given by 
$|\mu|\le 2t$. 
Otherwise, there is an energy gap and the Chern number
is zero.

\begin{figure}[t]
\begin{center}\leavevmode
\includegraphics[width=0.38\linewidth]{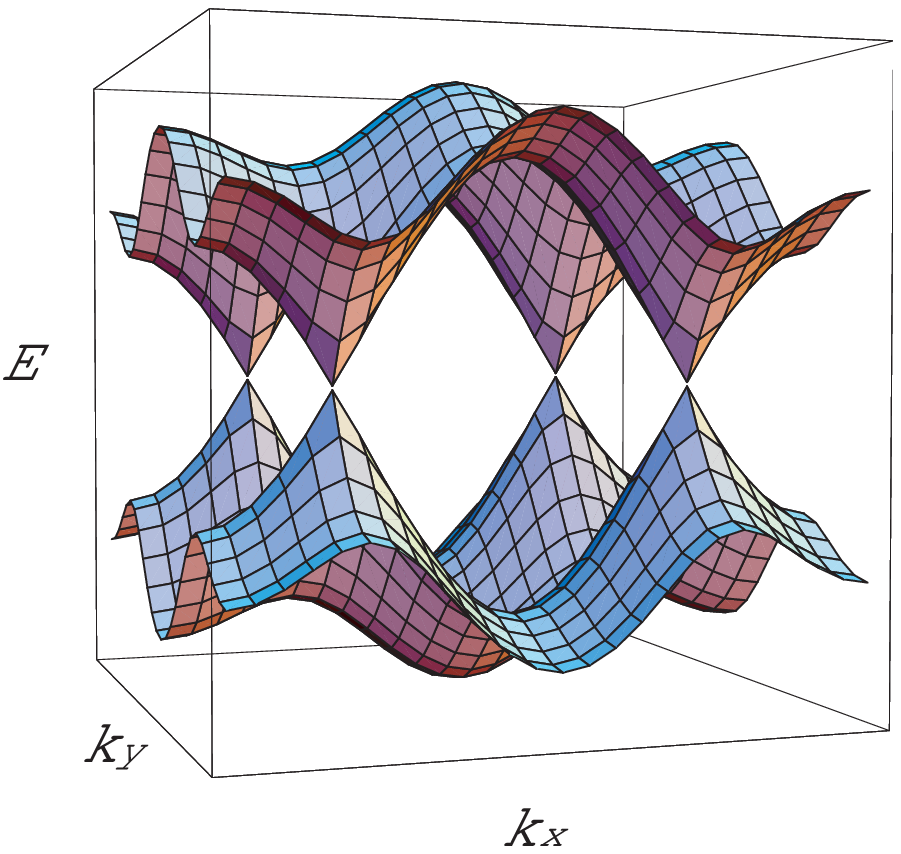}
\includegraphics[width=0.38\linewidth]{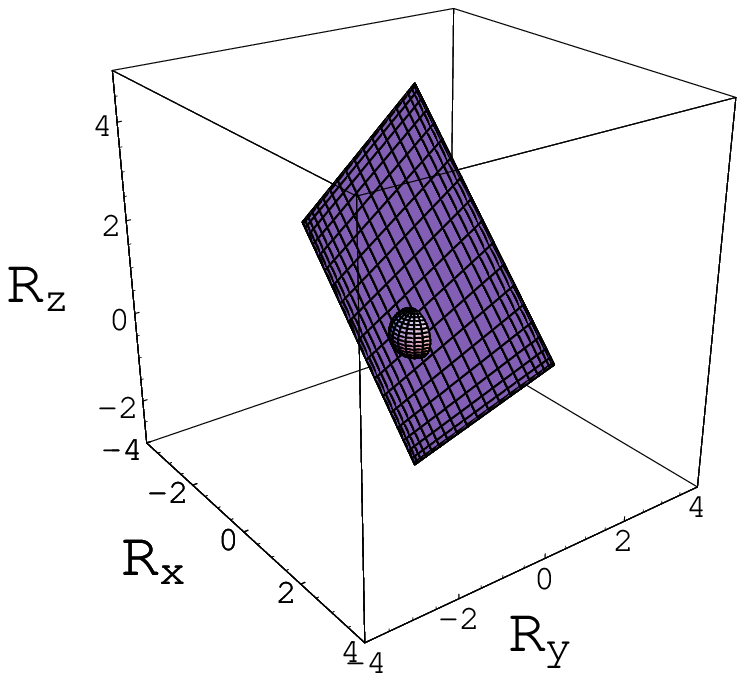}
\caption{ The energy dispersion and the mapped surface $R(T^2)$
with Dirac  monopole.
($t=\Delta_{x^2-y^2}=1,\mu=-t$ )} 
\label{figurename}\end{center}\end{figure}

Another interesting situation can be on the triangular lattice.
Starting from the t-J model on the triangular lattice which
is a typical correlated system with frustration, 
we get the B-dG equation on the triangular lattice. 
We take mean field ansatz, as the hopping are non zero for 
$ t_{i,i+\hat x}=t_{i+\hat x,i+\hat y}=t_{i,i+\hat y}=t$ and 
a possible order parameters for the 
superconductivity on the triangular lattice as
$
\Delta_{i,i+\hat x}=e^{i 2\pi/3} \Delta_{i+\hat x,i+\hat y}=e^{i
4\pi/3} \Delta_{i,i+\hat y}=\Delta
$.
Then $\epsilon(\mb{k} ) = -2t (\cos k_x+\cos k_y+\cos (k_x-k_y))$ 
and  $\Delta(\mb{k} ) = 2\Delta (\cos k_x+e^{i 2\pi/3}\cos k_y+e^{i
4\pi/3}\cos (k_x-k_y))$.
 It clearly shows that the covering degree of mapping and the
intersection number is 2 ( The Chern number is 2). 
It suggests there may exist a non trivial topological order on the 
possible 
superconductivity on the triangular lattice. 
(Detailed discussions  will be given elsewhere.)

\begin{figure}[btp]
\begin{center}\leavevmode
\includegraphics[width=1.1\linewidth]{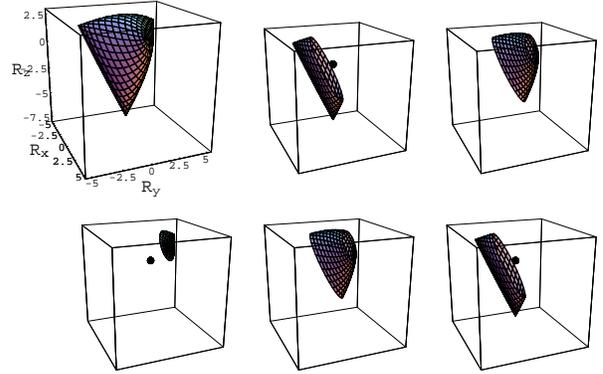}
\caption{
Mapped Brillouin zone $R(T^2)$ for the triangular lattice.
The monopole at the origin $O$ is also shown.
(a) is the total surface $R(T^2)$, $k_x\in {(}0,2\pi{]}$, $k_y\in {(}0,2\pi{]}$
and (b)-(f) are parts of the surface to show how it covers the origin.
$\Delta  _0=0,\ \Delta_{x}= \Delta_{y}= \Delta_{xy}=t,\ \mu=0$. 
}
\label{figurename}\end{center}\end{figure}

%
%



\begin{thebibliography}{99}

\bibitem{wen_topological}
X. G. Wen,Phys. \ Rev.\ B 40, 7387 (1989).



\bibitem{tknn}
D. J. Thouless, M. Kohmoto, P. Nightingale, M. den Nijs,
Phys.\ Rev.\ Lett. 49, 405 (1982),
M. Kohmoto, Ann.\  Phys.\  (N. Y. ) 160, 355 (1985).

\bibitem{yh}
Y. Hatsugai, J.\  Phys. \ C,\  Condens. \ Matter,
9, 2507, (1997), Phys.\ Rev. \  B48, 11851, (1993),
Phys.\ Rev. \ Lett. 71, 3697 ( 1993).

\bibitem{laughlin-dd}
R. B.Laughlin, Phys. \ Rev. \ Lett. 80, 5188 (1998)

\bibitem{volovik1}
G. E. Volovik,JETP Lett. 66, 493 (1997)

\bibitem{smf}
T. Senthil, J.B.Marston, M.P.A. Fisher,
Phys. \ Rev.\ B 60, 4245, (1999).

\bibitem{ym-yh} Y. Morita, Y. Hatsugai, Phys.\  Rev.\  B62, 99 (2000),
Phys.\ Rev.\ Lett. 86, 51 (2000)
 

\bibitem{yh-sr} Y. Hatsugai, S. Ryu, to appear in Phys. Rev. B
(2002).



\bibitem{berry}
T. T. Wu, C. N. Yang, Phys.\ Rev. \ D12, 3845 (1975),
M. V. Berry, Proc.\ R.\ Soc. 45, A392, (1984)





\end{thebibliography}

\end{document}